\documentclass{article}
\usepackage{ijcai26}

\usepackage{times}
\usepackage{soul}
\usepackage{url}
\usepackage[hidelinks]{hyperref}
\usepackage[utf8]{inputenc}
\usepackage[small]{caption}
\usepackage{graphicx}
\usepackage{amsmath}
\usepackage{amsthm}
\usepackage{booktabs}
\usepackage{algorithm}
\usepackage{algorithmic}
\usepackage{amssymb}
\usepackage{makecell}
\usepackage{authblk}
\usepackage[table]{xcolor}
\usepackage[switch]{lineno}

\title{COCO: Cognitive Operating System with Continuous Oversight for Multi-Agent Workflow Reliability}
\author[1]{Churong Liang}
\author[1]{Jinling Gan}
\author[1]{Kairan Hong}
\author[1]{Qiushi Tian}
\author[1]{Zongze Wu}
\author[1]{Runnan Li}
\affil[1]{Beijing University of Posts and Telecommunications}

\begin{document}

\maketitle

\begin{abstract}
A critical limitation in large-scale multi-agent systems is the cascading of errors. And without intermediate verification, downstream agents exacerbate upstream inaccuracies, resulting in significant quality degradation. To bridge this gap, we introduce \textbf{COCO} (\textbf{C}ognitive \textbf{O}perating System with \textbf{C}ontinuous \textbf{O}versight), a theoretically grounded framework for asynchronous self-monitoring and adaptive error correction in multi-agent systems. COCO reconciles the fundamental tension between quality assurance and computational efficiency via a novel decoupled architecture. This design isolates error detection from the critical execution path and incorporates an automated configuration engine to minimize deployment complexity. The framework relies on three algorithmic innovations to mitigate both systematic and stochastic errors: (1) a Contextual Rollback Mechanism that leverages execution history for informed state recovery rather than naive retries; (2) a Bidirectional Reflection Protocol to ensure convergence and prevent oscillatory control loops; and (3) a Heterogeneous Cross-Validation Mechanism that utilizes ensemble disagreement to identify bias and hallucinations. Extensive experiments on diverse benchmarks demonstrate that COCO delivers a 6.5\% average performance improvement. Notably, the framework achieves 95.1\% of large-model performance with a 30$\times$ parameter reduction, confirming the potential for efficient, high-reliability deployment, and establishing COCO as a practical, annotation-based solution for critical autonomous domains.

\end{abstract}

\section{Introduction}

The transition toward compositional intelligence via multi-agent systems has become a defining characteristic of modern AI~\cite{wu2023autogen}. By decomposing high-dimensional reasoning tasks into specialized, collaborating modules, these architectures achieve unprecedented scalability and adaptability~\cite{qian2025scaling}. However, this modularity introduces a critical structural vulnerability: the sequential interdependencies required for collaboration create pathways for error amplification. Consequently, minor upstream hallucinations can cascade through the workflow, catastrophically compromising global system integrity~\cite{cemri2025mast}.

\textbf{The Error Propagation Problem.} For a formalized multi-agent workflow $ W = \{A_1, A_2, \ldots, A_n\}$, where each agent $ A_i$ transforms an intermediate representation $ z_{i-1}$ into an output $ z_i = A_i(z_{i-1}; \theta_i)$. A critical vulnerability in this sequential dependency is the compounding nature of approximation errors. Small perturbations $\epsilon_i$ at early stages do not merely persist; they act as multiplicative factors in downstream deviations \cite{snowball2025,peng-etal-2025-stepwise}. This phenomenon is bounded by $ \mathbb{E}[\|z_n - z_n^*\|] \leq \prod_{i=1}^n (1 + \epsilon_i)$ where $z_n^*$ denotes the optimal output trajectory. We term this epistemic error cascading, a mechanism where stochastic noise accumulates exponentially with pipeline depth, inevitably destabilizing complex reasoning tasks ~\cite{caselli2015piling}.

\begin{figure}[t]
\centering
\includegraphics[width=0.9\columnwidth]{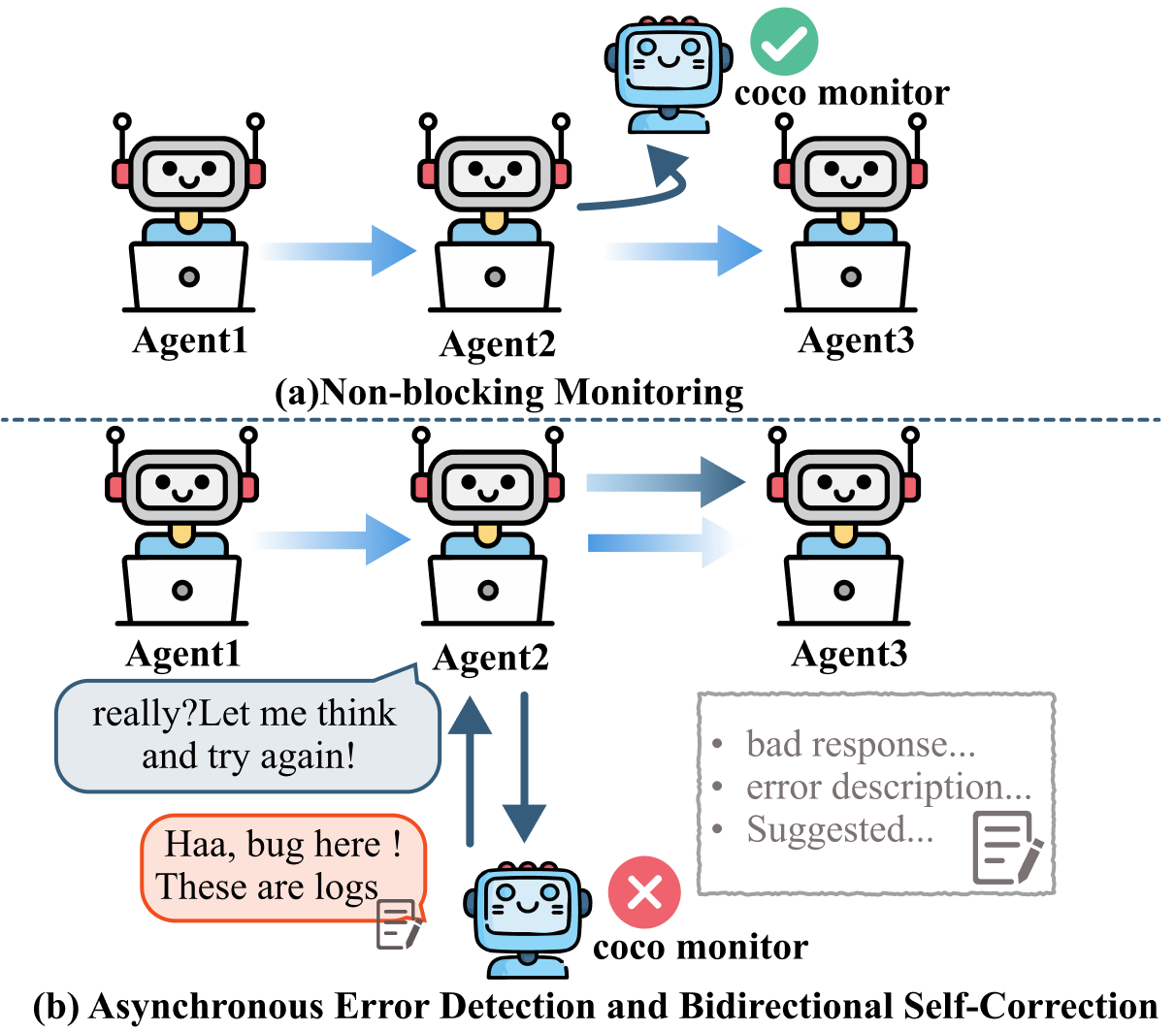}
\caption{By intercepting intermediate execution errors through asynchronous monitoring, COCO initiates a self-repair loop that prevents the downstream propagation of inaccuracies.}
\label{fig1}
\end{figure}

\textbf{Limitations of Existing Approaches.} Current fault-tolerance strategies fail to reconcile the tension between low-latency inference and autonomous self-repair. One category of solutions, including circuit-breaker patterns~\cite{zou2024circuit} and gatekeeping mechanisms like LLMGuard~\cite{llmguard2024}, excels at interception but suffers from semantic unawareness. These systems function as terminal blockers: they can halt pathological outputs but lack the contextual understanding required to initiate intelligent recovery. Conversely, synchronous validation frameworks such as SelfCheck and Self-Refine~\cite{selfcheckgpt2023,selfrefine2023} improve reasoning fidelity but introduce prohibitive latency overheads. By forcing sequential verification steps, they create execution bottlenecks that render them impractical for scalable, latency-sensitive workflows.

\textbf{COCO Framework.} To resolve the fundamental tension between workflow reliability and computational efficiency, we propose \textbf{COCO} (\textbf{C}ognitive \textbf{O}perating System with \textbf{C}ontinuous \textbf{O}versight). COCO departs from synchronous validation paradigms by adopting a decoupled asynchronous architecture. This design effectively isolates the Monitoring Engine from the critical execution path, ensuring $O(1)$ overhead relative to workflow complexity.

To address the specific failure modes of error amplification, oscillation, and bias, COCO orchestrates three synergistic mechanisms:
(1) The Contextual Rollback Mechanism (\textbf{CRM}), which replaces naive retries with stateful recovery. By preserving execution snapshots ($\Psi_i$) and injecting diagnostic patches ($\Delta_i$), CRM enables informed re-computation that halts epistemic error cascading.
(2) The Bidirectional Reflection Protocol (\textbf{BRP}), which mitigates control-loop instability. By establishing a symmetric consensus protocol between execution and monitoring nodes, BRP prevents correction oscillations and ensures convergence.
(3) Heterogeneous Cross-Validation (\textbf{HCV}), which targets systematic bias. By leveraging information-theoretic disagreement across diverse model architectures, HCV detects subtle hallucinations that escape homogenous self-checks. 

The main contributions are summarized as follows: 
\begin{itemize} 
\item Decoupled Asynchronous Oversight: We redefine error handling in multi-agent systems via a non-intrusive architecture that isolates oversight from the execution path. Unlike synchronous validators that create linear latency bottlenecks, COCO maintains comprehensive error detection with minimal computational overhead.
\item Algorithmic Novelty: We introduce a tripartite error-correction stack comprising CRM for history-aware state restoration instead of naive retries, BRP for achieving stable actor-critic consensus, and HCV for bias-resistant detection through heterogeneous ensemble disagreement analysis.
\item Production-Ready Integration: COCO provides a complete annotation-based platform that seamlessly integrates with native checkpointing mechanisms, supported by an automated configuration engine that optimizes monitoring strategies via empirical sampling. 
\item State-of-the-Art Efficiency: Extensive benchmarking demonstrates that COCO bridges 53.9\% of the performance gap between compact and Titan-scale models. Crucially, it achieves 95.1\% of large-model performance with a 30$\times$ parameter reduction, delivering a consistent 6.5\% improvement over strong baselines. 
\end{itemize}

\section{Related Work}
\textbf{From Isolated Generation to Multi-Agent Workflows.} The trajectory of large language models (LLMs) has shifted from isolated generation to the paradigm of compositional intelligence. Early methodologies enhanced individual reasoning capabilities through Chain-of-Thought (CoT) and ReAct paradigms, enabling models to decompose complex queries into executable steps~\cite{cot2022,react2023}.
Recent frameworks have scaled this capability into Multi-Agent Systems (MAS), where heterogeneous models are auto-orchestrated into graph-based workflows~\cite{aflow2025}. Architectures such as AutoGen~\cite{autogen2023} and AgentVerse~\cite{agentverse2024} distribute subtasks across specialized roles, while structured mechanisms like dynamic agent pools~\cite{collab2024} and fixed-role pipelines~\cite{mapcoder2024} optimize inference efficiency. However, this transition from single-turn inference to sequential, interdependent collaboration inherently increases the depth of the execution graph. As formalized before, this expanded interaction surface creates the structural conditions necessary for epistemic error cascading.

\textbf{Dynamics of Error Propagation.} Recent empirical studies validate the "snowballing" phenomenon in epistemic error cascading. When a high-error-rate agent appears in a graph-structured workflow, global accuracy degrades exponentially with interaction rounds. In linear structures, this decline can exceed 30\%~\cite{faultyagents2025}. The BEHAVIORCHAIN benchmark similarly observes that initial approximation errors are amplified layer-by-layer during multi-round cooperation~\cite{behaviorchain2025}.

This phenomenon has been analyzed through information-theoretic frameworks~\cite{snowball2025} and uncertainty decomposition (UProp)~\cite{uprop2025}, which reveal that early-stage hallucinations function as multiplicative noise. Crucially, counterfactual analysis suggests that precisely repairing key nodes can halve system failure rates~\cite{critical2025}, providing the theoretical motivation for the Contextual Rollback Mechanism employed in the proposed framework.

\textbf{Auditing and Guardrail Paradigms.} Current approaches to mitigating these errors fall into two primary categories, each with distinct limitations regarding the latency-reliability trade-off:

\textit{1) Synchronous Self-Correction:} Methods such as SelfCheckGPT~\cite{selfcheckgpt2023} and Self-Refine~\cite{selfrefine2023} employ iterative \textit{generate–critique–rewrite} loops to reduce hallucinations. While effective—delivering average improvements of 20\%—these approaches are inherently synchronous, introducing linear latency penalties that render them impractical for real-time applications.

\textit{2) Runtime Guardrails:} To ensure safety without the latency of self-correction, systems employ external validation. Strategies include crowdsourcing-style parallel voting~\cite{llmchains2024} and deterministic gatekeepers like LLMGuard~\cite{llmguard2024}. While these mechanisms effectively intercept pathological outputs, they function as terminal blockers with limited semantic understanding, often lacking the context required for the intelligent state recovery that the proposed framework provides.

\begin{figure*}[th]
\centering
\includegraphics[width=0.95\textwidth]{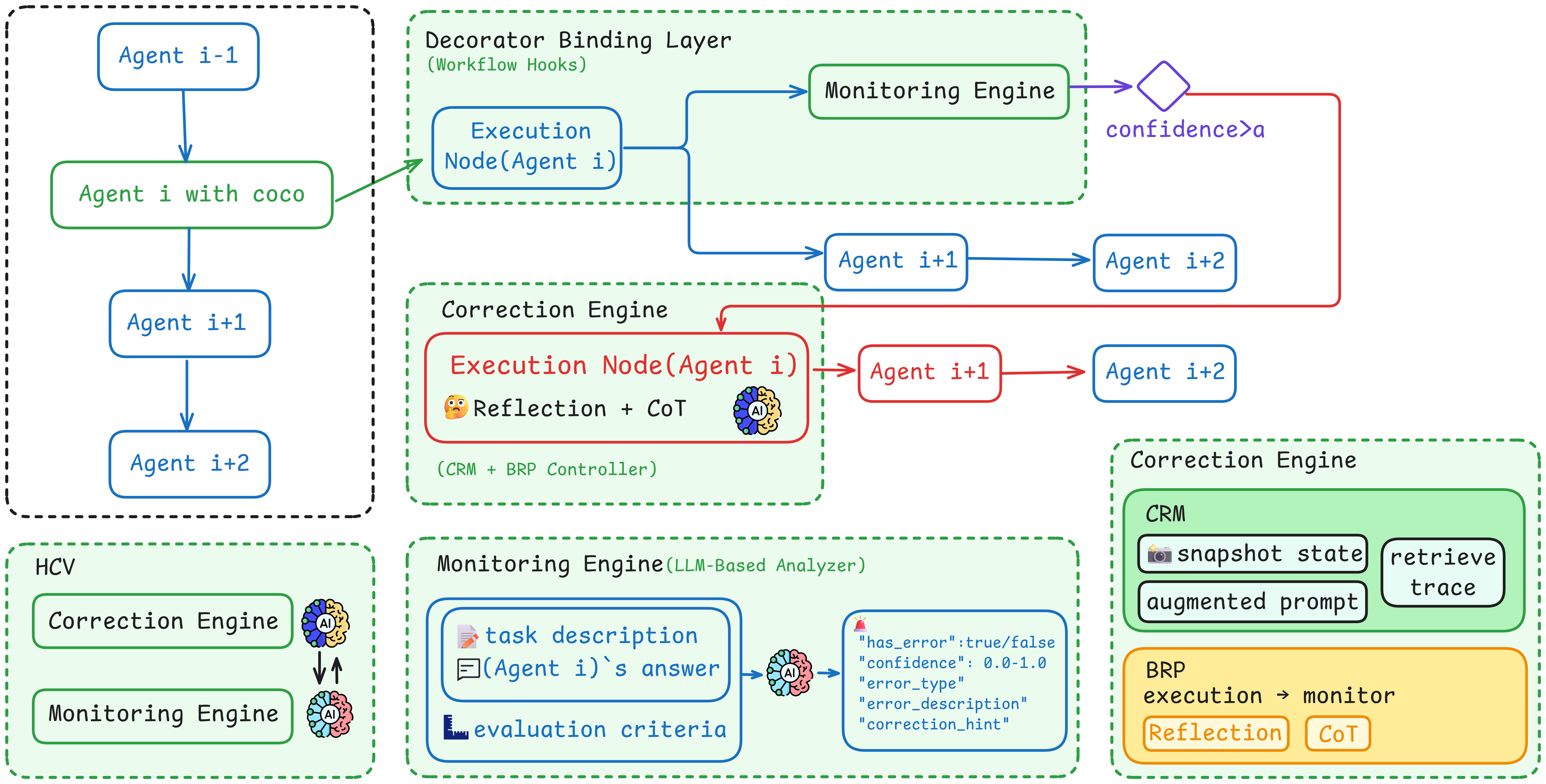}
\caption{Architectural overview of the COCO framework. The system is composed of three decoupled layers: (1) The \textbf{Decorator Binding Layer} facilitates non-intrusive, asynchronous oversight via agent-level hooks; (2) The \textbf{Monitoring Engine} leverages LLMs to synthesize structured diagnostic reports from execution outputs; and (3) The \textbf{Correction Engine} orchestrates state restoration, integrating CRM for contextual rollback with BRP and HCV to guarantee correction stability.}

\label{fig2}
\end{figure*}

\section{Methodology}

Multi-agent workflows are inherently vulnerable to error cascading, yet current mitigation strategies face a critical dichotomy: synchronous validation imposes linear latency penalties ($O(n)$), while stateless retries lack the semantic context for recovery. COCO reconciles this tension by leveraging native graph-based checkpointing to implement asynchronous decoupled monitoring. Unlike invasive state-management solutions, COCO integrates seamlessly with existing frameworks, utilizing parallel oversight to enforce quality guarantees with constant $\mathcal{O}(1)$ overhead relative to workflow depth.

\subsection{Problem Definition}

Multi-agent workflows suffer from four critical failure modes that compromise reliability:

\textbf{Error Propagation and Amplification:} In sequential agent chains, errors from upstream agents cascade into downstream processes, often amplifying in severity. Without intervention, minor initial errors can render entire workflow
outputs unusable.

\textbf{Context Loss During Recovery:} Traditional retry mechanisms discard the execution state $S_i$ upon failure. This loss of context causes agents to repeat identical mistakes or introduce new errors during recovery attempts due to a lack of informed diagnostic history.

\textbf{Correction Oscillations:} Naive error correction systems often generate conflicting fixes that prevent convergence. Without a symmetric evaluation protocol like BRP, competing correction mechanisms may introduce contradictory changes that fail to stabilize the reasoning path.

\textbf{Performance-Quality Tradeoff:} Conventional validation approaches introduce significant computational overhead that scales linearly, $O(n)$, with workflow depth. This forces system designers to compromise between thorough validation and real-time performance.

\subsection{Error Correction Components}
COCO leverages the complementary strengths of its three core mechanisms to address these challenges: Contextual Rollback Mechanism (CRM), Bidirectional Reflection Protocol (BRP), and Heterogeneous Cross-Validation (HCV).

\subsubsection{Contextual Rollback Mechanism}
To mitigate the stochastic degeneracy inherent in stateless retry heuristics, CRM implements a history-aware state recovery protocol. Unlike naive approaches that discard the execution context upon failure, CRM persists a structured checkpoint $\Psi_i = \langle s_i, P_i \rangle$ for each agent $A_i$, comprising the pre-execution state $s_i$ and the original instruction set $P_i$.

\textbf{Trigger and Formulation.} Upon the detection of a high-confidence failure—formalized as the condition $\mathbb{I}(h_i = 1) \land (c_i \geq \alpha)$—CRM interrupts the downstream propagation and reverts the workflow to state $s_i$. Crucially, it constructs a Correction Patch $\Delta_i$ to condition the re-generation:
\begin{equation}
\Delta_i = \langle \delta_i, \gamma_i, z_i^{fail} \rangle
\end{equation}
where $\delta_i$ represents the diagnostic rationale, $\gamma_i$ denotes the corrective constraints, and $z_i^{fail}$ serves as a negative sample.

\textbf{Informed Re-computation.} The re-execution mechanism is modeled as a counterfactual generation process:
\begin{equation}
z_i' = A_i(s_i, P_i^ ; \theta_i) \quad 
\text{s.t.} \quad P_i^ = \mathcal{F}{aug}(P_i, \Delta_i)
\end{equation}Here, the augmentation function $\mathcal{F}{aug}$ integrates the patch $\Delta_i$ into the prompt manifold. By explicitly incorporating the failure mode $z_i^{fail}$ and the diagnostic $\delta_i$, CRM effectively modifies the optimization landscape, penalizing the previously taken path. This transforms the recovery process from a random resampling event into a guided local search, ensuring that $z_i'$ diverges from locally convergent failure modes and effectively curbing epistemic error cascading.

\subsubsection{Bidirectional Reflection Protocol}

To address the limitations of unidirectional validation, BRP implements a Dual-Phase Rectification Topology. Unlike iterative feedback systems that introduce unbounded latency, BRP operates as a strictly bounded, single-turn consensus mechanism. This design guarantees system stability by enforcing a hard break in the feedback loop, relying on the depth of reasoning (Chain-of-Thought) rather than the frequency of iteration to ensure quality. The protocol executes in two deterministic phases:

\textbf{Phase 1: Diagnostic Projection (Monitor Side).}
Upon detecting an anomaly, the monitoring engine $\mathcal{M}$ does not merely reject the output but projects a structured diagnostic vector $\mathbf{R}_i$ into the agent's context window. Formally, given the initial faulty output $z_i^{fail}$ and agent state $s_i$, the monitor computes:
\begin{equation}
\mathbf{R}_i = \mathcal{M}(z_i^{fail}, \Psi_i) = \langle h_i, c_i, \tau_i, \delta_i, \gamma_i \rangle
\end{equation}

where $\Psi_i$ denotes structured execution snapshots, $h_i$ indicates the presence of errors, $c_i$ means error type, $\tau_i$ identifies the error category, $\delta_i$ provides the diagnostic rationale (the "Why"), and $\gamma_i$ offers directive constraints (the "How") to guide the correction.

\textbf{Phase 2: Reflective Reconstruction (Agent Side).}
Instead of simply retrying, the execution agent $A_i$ enters a Reflective Chain-of-Thought (CoT) mode. It consumes $\mathbf{R}_i$ to strictly condition its re-generation process. The corrected output $z_i'$ is generated as:
\begin{equation}
z_i' = A_i(s_i, z_i^{fail}, \mathbf{R}i; \theta{cot})
\end{equation}
In this phase, the agent explicitly articulates an internal reasoning trace that aligns its intent with the constraints provided in $\gamma_i$ before producing the final result.

\textbf{Stability and Overhead Analysis.}
This rectification architecture guarantees both structural stability and computational efficiency by enforcing a strict Directed Acyclic Graph (DAG) topology in the correction logic. By structurally severing the feedback edge after the reconstruction phase, BRP inherently eliminates the possibility of control-loop oscillation while ensuring bounded overhead with a deterministic $O(1)$ inference cost, thereby preventing the latency spikes characteristic of indeterminate iterative refinement.

\subsubsection{Heterogeneous Cross-Validation}

In multi-agent collaboration, a single model often struggles to identify flaws in its own reasoning, making it prone to logical loops or cognitive biases. To enhance the system's self-auditing capability, COCO introduces the Heterogeneous Cross-Validation (HCV) mechanism.

HCV establishes a cross-architecture feedback pathway between execution and monitoring modules by: Employing fundamentally different model architectures for execution and monitoring, Leveraging the distinct "thinking patterns" of different models to detect subtle errors, Creating diversity in error detection capabilities that compensates for blind spots in any single model, Enabling bias-resistant detection through ensemble disagreement analysis.

By utilizing complementary model architectures with different inductive biases, HCV enables the monitoring system to detect subtle errors that the execution model might consistently overlook, thereby improving the system's error correction capability and overall robustness.

\begin{algorithm}[t]
    \caption{Asynchronous Oversight and Contextual Recovery via COCO}
    \label{alg:coco_algo}
    \textbf{Input}: Workflow $\mathcal{G}(\mathcal{V}, \mathcal{E})$, Input query $x$, Configuration $\mathcal{D}$ \\
    \textbf{Output}: Final verified output $z_n$
    \begin{algorithmic}[1]
        \STATE \textbf{Phase 1: Configuration Overlay Initialization}
        \IF{$\mathcal{D}$ is NULL}
            \STATE $\mathcal{D} \leftarrow \text{SampleAndTest}(\mathcal{G}, \text{tasks})$ \COMMENT{Identify high-risk nodes by testing}
        \ENDIF
        \STATE $\tilde{\mathcal{V}} \leftarrow \text{LoadMonitoringNodes}(\mathcal{D})$ \COMMENT{Load Monitor Config}
        
        \STATE \textbf{Phase 2: Execution \& Asynchronous Monitoring}
        \FOR{each agent node $A_i \in \text{Path}(\mathcal{G})$}
            \STATE $z_i \leftarrow A_i(s_i, P_i; \theta_{exe})$ \COMMENT{Execute original agent task}
            \IF{$A_i \in \tilde{\mathcal{V}}$}
                \STATE \textbf{Async Trigger:}\COMMENT{Spawn parallel monitor thread}
                \STATE $\mathbf{R}_i = \langle h_i, c_i, \tau_i, \delta_i, \gamma_i \rangle \leftarrow \mathcal{M}(z_i, \Psi_i)$ \COMMENT{Generate diagnostics}
                
                \IF{$h_i$ is True \textbf{AND} $c_i > \alpha$}
                    \STATE \textbf{Phase 3: Context-Aware Recovery (CRM + BRP)}
                    \STATE $\Delta_i \leftarrow \text{GenPatch}( \Psi_i, \mathbf{R}_i)$ \COMMENT{Patch Generation via context and diagnostics}
                    \STATE $P_i^* \leftarrow \mathcal{F}_{aug}(P_i, \Delta_i)$ \COMMENT{Prompt augmentation via the patch}
                    \STATE $z_i' \leftarrow A_i(s_i, P_i^*)$ \COMMENT{Reflective correction via BRP and CoT}
                    \STATE \textbf{Update} $z_i \leftarrow z_i'$ and re-route downstream execution
                \ENDIF
            \ENDIF
        \ENDFOR
        \STATE \textbf{return} $z_n$
    \end{algorithmic}
\end{algorithm}

\begin{figure*}[t]
\centering
\includegraphics[width=0.95\textwidth]{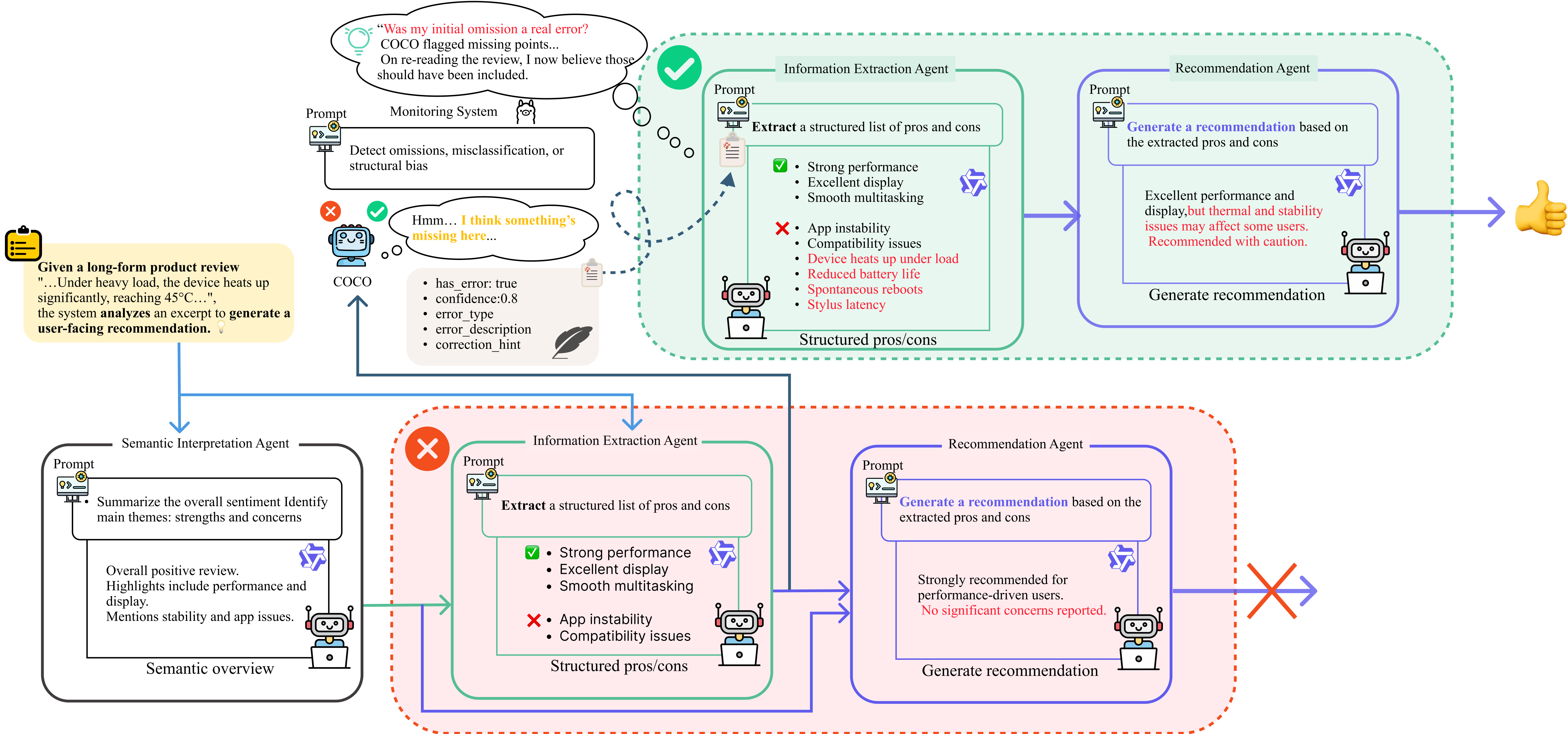}
\caption{Case study of a COCO-integrated workflow for generating recommendations from product reviews. COCO monitors node2 asynchronously; in the lower path, the workflow continues before correction occurs. In the upper path, COCO detects an omission and rolls back node2, producing an improved result.}
\label{fig3}
\end{figure*}

\subsection{Overall Architecture}
COCO implements a three-layer decoupled architecture that achieves intelligent monitoring and automatic correction through seamless integration with multi-agent workflows.

\textbf{Monitor Binding Layer}: It implements loose coupling by defining a set $\tilde{\mathcal{V}} \subseteq \mathcal{V}$ through non-invasive decorators. Monitored execution nodes $A_i$ complete tasks $z_i$ independently and trigger monitors $\mathcal{M}$ asynchronously, ensuring $O(1)$ overhead without blocking execution path. This mechanism supports dynamic configuration of strategies and ensures state isolation for reliable multi-instance coordination.

\textbf{Monitoring Engine}: The engine serves as an LLM-based analyzer $\mathcal{M}(z_i, \Psi_i; \theta_{mon})$ that performs multi-dimensional assessments covering logical consistency, format compliance, and content completeness. It transforms qualitative judgments into a structured diagnostic report $\mathbf{R}_i = \langle h_i, c_i, \tau_i, \delta_i, \gamma_i \rangle$, yielding quantitative confidence metrics $c_i$ used for threshold-based intervention.

\textbf{Correction Engine}: This engine manages workflow state transitions by orchestrating the CRM and BRP. It preserves structured execution snapshots $\Psi_i$ and integrates diagnostic information $\delta_i$ with historical context to generate an augmented prompt $P_i^* = \mathcal{F}_{aug}(P_i, \Delta_i)$. The correction process utilizes the BRP to guide the model through reflective re-computation $z_i'$, ensuring convergence to valid results.

The architecture supports four error categories $\tau_i \in \{\text{logic, format, content, systematic}\}$. COCO achieves seamless integration through annotation-based monitoring declarations that require minimal changes.

\begin{table*}[t]
  \centering
  \footnotesize %
  \setlength{\tabcolsep}{4pt} %
  \begin{tabular}{l r r r r r r} %
    \toprule
    \textbf{Task Metric} & \textbf{\makecell[r]{GSM-\\Hard}} & \textbf{\makecell[r]{MMLU-\\Pro}} & \textbf{MBPP} & \textbf{\makecell[r]{CommonGen-\\Hard}} & \textbf{AVG.} & \textbf{\makecell[r]{Improvement \\Overhead}} \\
    \midrule
    \rowcolor[HTML]{F0F0F0} \multicolumn{7}{l}{\textbf{Reference Models}} \\
    Qwen3-235B-A22B & \textbf{0.7210} & \textbf{0.7247} & \textbf{0.8120} & \textbf{0.8712} & \textbf{0.7822} & -- \\
    Qwen3-8B        & 0.6808          & 0.5885          & 0.6600          & 0.8115          & 0.6852          & -- \\
    Llama-3.1-8B    & 0.3669          & 0.4803          & 0.6020          & 0.7700          & 0.5548          & -- \\
    \addlinespace[3pt] 

    \rowcolor[HTML]{F0F0F0} \multicolumn{7}{l}{\textbf{Multi-Agent Framework Baseline}} \\
    Aflow-Qwen3-8B    & 0.6399 & 0.6656 & 0.6500 & 0.8389 & 0.6986 & -- \\
    Aflow-Llama3.1-8B & 0.5648 & 0.4514 & 0.5580 & 0.7492 & 0.5809 & -- \\
    \addlinespace[3pt]

    \rowcolor[HTML]{F0F0F0} \multicolumn{7}{l}{\textbf{COCO Framework (Proposed)}} \\
    COCO Qwen3-8B with coco(Llama-3.1-8B) & \underline{0.6989} & \underline{0.6869} & 0.7440 & 0.8450 & \underline{0.7437} & +6.50\% \\
    COCO Qwen3-8B with coco(Qwen3-8B)  & 0.6875 & 0.6660 & \underline{0.7660} & \underline{0.8477} & 0.7418 & +6.20\% \\
    COCO Llama-3.1-8B with coco(Qwen3-8B)  & 0.6023 & 0.5344 & 0.6040 & 0.8028 & 0.6359 & +9.50\% \\
    COCO Llama-3.1-8B with coco(Llama-3.1-8B) & 0.5201 & 0.4562 & 0.5540 & 0.8081 & 0.5846 & +0.63\% \\
    \bottomrule
  \end{tabular}
  \caption{Performance benchmarking across diverse datasets. We compare COCO against standard reference models and the state-of-the-art AFlow multi-agent baseline. All improvement metrics report gains relative to the corresponding AFlow workflow.}
  \label{tab:table1}
\end{table*}

\begin{table}[t]
\centering
\footnotesize 
\begin{tabular}{lrr}
\toprule
\textbf{System Configuration} & \textbf{GSM-Hard} & \textbf{Avg. Latency (s)} \\ 
\midrule
AFlow-Qwen3-8B     & 0.6399          & 15.6 \\
COCO(Qwen3-8B)     & 0.6875          & 22.7 \\
COCO(Llama-3.1-8B) & \textbf{0.6989} & \textbf{15.0} \\ 
\bottomrule
\end{tabular}

\caption{Average latency comparison with Qwen3-8B backbone. COCO's heterogeneous configuration maintains parity with the AFlow's efficiency while significantly enhancing task quality.}
\label{tab:table2}
\end{table}

\section{Experiments and Discussion}

\subsection{Datasets and Metrics}
To assess error correction robustness, we benchmark COCO on diverse, logically demanding tasks covering both closed- and open-domain settings:

\textbf{Closed-domain Tasks:} These tasks require precise, objective reasoning with unambiguous correct answers, making them ideal for evaluating core error detection accuracy and correction effectiveness. \textbf{GSM-Hard}\cite{gsmhard2025} features arithmetic problems with complex multi-step calculations and large numerical values, challenging the framework's ability to detect and correct mathematical reasoning errors while maintaining computational accuracy. \textbf{MMLU-Pro}\cite{mmlupro2024} provides a comprehensive benchmark spanning diverse knowledge domains through multiple-choice questions, testing both factual recall accuracy and logical deduction correctness. Both benchmarks use accuracy as the evaluation metric, enabling precise quantification of error correction improvements.

\textbf{Open-domain Tasks:} These tasks are inherently creative and open-ended, requiring multi-dimensional qualitative evaluation of error correction in complex generative workflows. \textbf{MBPP}\cite{mbpp2021} tasks agents with generating functionally correct code solutions, demanding proficiency in syntax error detection, logical flow correction, and semantic consistency validation. The evaluation combines code executability, test case passing rates, and algorithmic correctness, reflecting practical software development error patterns. \textbf{CommonGen-Hard}\cite{selfrefine2023} challenges agents to generate coherent sentences connecting disparate concepts, testing error correction in commonsense reasoning, contextual understanding, and creative expression. Evaluation incorporates grammar accuracy, logical consistency, concept coverage, and semantic coherence, providing comprehensive assessment of generative quality.

\subsection{Implementation Details}
The COCO framework implementation integrates four key architectural innovations through systematic engineering approaches that maintain theoretical guarantees while enabling practical deployment.

\textbf{AFlow Iterative Optimization Framework.} We employed the AFlow framework to automatically construct and optimize multi-agent workflows through Monte Carlo Tree Search-based exploration. This process generates DAG structures with performance-guided agent dependencies. For fair comparative evaluation, we augmented baseline configurations with analysis nodes (the first node) to align computational complexity with COCO monitoring overhead.

\textbf{COCO Decorator Architecture.} We implemented monitoring via a non-intrusive decorator pattern, enabling seamless integration without altering core workflow logic. To balance flexibility and automation, COCO employs a hybrid configuration strategy: high-risk nodes are identified either through empirical manual selection or an automated SampleAndTest protocol. This process generates an optimized overlay $\mathcal{D}$ that populates the monitoring set $\tilde{\mathcal{V}}$ for oversight.

\textbf{Heterogeneous Multi-Agent vLLM\cite{vllm2023} Parallel Processing.} Our local model deployment utilizes vLLM's asynchronous framework to support concurrent multi-agent processing across heterogeneous architectures:
\begin{itemize}
\item Small-scale models (Qwen3-8B\cite{qwen3report2025}, Llama-3.1-8B-Instruct-Turbo\cite{llama3_2024}) run through local vLLM instances for high-throughput parallel execution.
\item Large-scale models (Qwen3-235B-A22B\cite{qwen3report2025}) are accessed through API interfaces.
\item The architecture integrates diverse model families through standardized interfaces while maintaining uniform performance characteristics.
\end{itemize}

\subsection{Experimental Results}
The quantitative evaluation results in Table~\ref{tab:table1} state the effectiveness of COCO to: i) achieve consistent performance improvements across all model combinations, with relative gains ranging from 0.63\% to 9.5\% over AFlow baselines, ii) attain 95.1\% of the state-of-the-art Qwen3-235B-A22B performance using only 8B parameters—a 30× reduction in model size while maintaining near-equivalent quality, and iii) bridge 53.9\%  of the performance gap between AFlow baselines and the large-scale model upper bound through effective error correction. All these aspects reflect COCO's effectiveness as an error correction framework in promoting downstream multi-agent workflow reliability.

\textbf{Computational Efficiency Analysis.} Table~\ref{tab:table2} presents the computational efficiency of different COCO system configurations. Specifically, \textbf{COCO Qwen3-8B with coco(Llama-3.1-8B)} achieves an average latency of \textbf{15.0 seconds}, closely matching the baseline \textbf{AFlow-Qwen3-8B} at \textbf{15.6 seconds}. This confirms that COCO’s error monitoring does not introduce significant computational overhead when deployed in a heterogeneous setting. In contrast, the homogeneous configuration \textbf{COCO Qwen3-8B with coco(Qwen3-8B)} exhibits a higher latency of \textbf{22.7 seconds}, which we attribute to identical model execution paths causing redundant inference patterns. These results highlight the architectural advantage of leveraging heterogeneity (e.g., using Qwen3-8B for primary generation and Llama-3.1-8B for monitoring) to improve efficiency without sacrificing performance.

\begin{table}[t]
\centering
\footnotesize %
\begin{tabular}{lrrr} %
\toprule
\textbf{Configuration} & \textbf{GSM-Hard} & \textbf{MBPP} & \textbf{Average} \\
\midrule
w/o Reflection w/o CoT & 0.6880 & 0.6920 & 0.6900 \\
w/ Reflection w/o CoT  & \underline{0.6860} & \underline{0.6800} & \underline{0.6830} \\
w/o Reflection w/ CoT  & \textbf{0.7000} & 0.7300 & 0.7150 \\
w/ Reflection w/ CoT   & 0.6989 & \textbf{0.7440} & \textbf{0.7215} \\
\bottomrule
\end{tabular}
\caption{Ablation study examining individual and combined contributions of COCO's abilities. Results demonstrate synergistic effects between Bidirectional Reflection Protocol and CoT reasoning.}
\label{tab:table3}
\end{table}

\subsection{Ablation Study Results}

Table~\ref{tab:table3} presents comprehensive component analysis examining individual contributions of COCO's algorithmic innovations, revealing fundamental insights into the interplay between monitoring protocols and reasoning capabilities.

\textbf{Component Analysis.} The complete COCO configuration (w/ Reflection w/ CoT) achieves optimal performance at 0.7215 average, representing a 4.6\% improvement over the baseline configuration (w/o Reflection w/o CoT: 0.6900). This validates COCO's core hypothesis: systematic error correction through the Contextual Rollback Mechanism with complete Bidirectional Reflection Protocol produces multiplicative benefits. The ablation results reveal an asymmetry between COCO's core components—Chain-of-Thought without BRP (0.7150) significantly outperforms Bidirectional Reflection Protocol without CoT (0.6830) by 4.7\%, demonstrating that enhanced reasoning capabilities provide substantial value when integrated into the CRM framework. The performance progression from baseline (0.6900) to CoT integration (0.7150) to full system (0.7215) shows that CoT integration provides substantial improvement (+3.6\%), while the complete BRP implementation adds additional refinement (+0.9\%) when combined with CRM.

\subsection{Discussion}

\textbf{Reflection-only Performance Analysis.} The w/ Reflection w/o CoT configuration shows decreased performance (0.6830, -1.0\% from baseline), which can be attributed to the technical implementation of the rollback mechanism. When COCO activates error correction through contextual rollback, it incorporates previous erroneous outputs into the extended context window to guide the reflection process. Without Chain-of-Thought reasoning capabilities, this expanded context—containing error information—becomes prone to hallucination and confusion, as the model lacks the structured reasoning framework necessary to effectively process and learn from the error feedback. This technical limitation validates the necessity of the complete BRP design (Reflection + CoT) within the CRM framework, where CoT provides the essential reasoning structure that enables effective utilization of the contextual rollback information.

\section{Limitations and Future Work}

\subsection{Technical Limitations}
COCO exhibits several critical failure modes that limit its effectiveness in complex multi-agent environments. At the architectural level, the framework faces four challenges: (1) \textit{Cascade failures} when multiple agents fail simultaneously, overwhelming correction capacity; (2) \textit{Detection blind spots} where subtle systematic errors evade all three monitoring mechanisms; and (3) \textit{Resource exhaustion} when error rates exceed assumptions, causing queue overflow and monitoring delays. Beyond these operational constraints, COCO’s theoretical foundations also show limitations, as the framework assumes independent errors across agents while practical workflows often exhibit correlated failures due to shared dependencies, input distributions, or model architectures. This correlation violates the independence assumption underlying convergence guarantees, potentially leading to underestimated error propagation and inadequate correction capacity. Additionally, COCO’s post-execution detection requires workflow endpoints to await completion, and upon error detection, the framework restarts from the erroneous node while the original workflow continues, creating resource waste and added complexity.

\subsection{Ethical Considerations}

While error correction aims to detect systematic biases, mechanisms may inadvertently amplify them if error patterns align with protected attributes. Ongoing bias monitoring is vital for responsible deployment. COCO's complex logic can reduce interpretability, making it harder for users to understand why outputs change. Developing explainable correction methods remains challenging. Moreover, the interplay of the Contextual Rollback Mechanism, Bidirectional Reflection Protocol, and Heterogeneous Cross-Validation creates layered pathways that obscure causal links between errors and corrections. This opacity weakens user trust in high-stakes settings where understanding system behavior is critical for validation and accountability.

\subsection{Future Research Directions}

Future research should focus on developing meta-monitoring capabilities to supervise and evaluate COCO's own performance, creating a hierarchical oversight system. This includes developing metrics to assess the effectiveness of error detection, measuring the quality of rollback decisions, and monitoring the convergence behavior of correction mechanisms. Such self-supervision capabilities would enable COCO to adapt its monitoring strategies based on performance feedback and domain-specific error patterns, ultimately improving the reliability and effectiveness of the oversight framework itself.

\section{Conclusion}
COCO introduces a novel framework for detecting andcorrecting errors in multi-agent AI systems through asynchronous decoupled monitoring. By integrating Contextual Rollback Mechanism, Bidirectional Reflection Protocol, and
Heterogeneous Cross-Validation, COCO addresses the fundamental challenge of error propagation while maintaining O(1) overhead relative to workflow complexity.

Experimental on diverse benchmarks show COCO’s effectiveness, delivering 6.5\% average performance improvements and achieving 95.1\% of large-model performance with 30× parameter reduction. The ablation studies confirm
the synergistic relationship between reflection-based monitoring and chain-of-thought reasoning in optimal error correction. 

Beyond performance, COCO enables reliable deployment of modular AI systems in critical domains where error cascades could cause severe failures. Its annotation-based integration lowers implementation barriers, Furthermore, an automated configuration engine optimizes monitoring strategies via empirical sampling to minimize deployment complexity, making COCO a practical solution for improving multi-agent system reliability in production.

\clearpage
\bibliographystyle{named}
\bibliography{ijcai26}

\begin{thebibliography}{}

\bibitem[\protect\citeauthoryear{Caselli \bgroup \em et al.\egroup }{2015}]{caselli2015piling}
Tommaso Caselli, Piek Vossen, et~al.
\newblock When it's all piling up: investigating error propagation in an {NLP} pipeline.
\newblock In {\em Proceedings of the Workshop on {NLP} Applications: Completing the Puzzle, {WNACP} 2015, co-located with the 20th International Conference on Applications of Natural Language to Information Systems {(NLDB} 2015), Passau, Germany, June 17-19, 2015}, volume 1386 of {\em {CEUR} Workshop Proceedings}. CEUR-WS.org, 2015.

\bibitem[\protect\citeauthoryear{Cemri \bgroup \em et al.\egroup }{2025}]{cemri2025mast}
Mert Cemri, Melissa~Z Pan, et~al.
\newblock Why do multi-agent {LLM} systems fail?
\newblock In {\em The Thirty-ninth Annual Conference on Neural Information Processing Systems Datasets and Benchmarks Track}, 2025.

\bibitem[\protect\citeauthoryear{Chakraborty \bgroup \em et al.\egroup }{2025}]{collab2024}
Souradip Chakraborty, Sujay Bhatt, Udari~Madhushani Sehwag, et~al.
\newblock Collab: Controlled decoding using mixture of agents for {LLM} alignment.
\newblock In {\em The Thirteenth International Conference on Learning Representations}, 2025.

\bibitem[\protect\citeauthoryear{Chen \bgroup \em et al.\egroup }{2021}]{mbpp2021}
Mark Chen, Jerry Tworek, et~al.
\newblock Evaluating large language models trained on code, 2021.

\bibitem[\protect\citeauthoryear{Chen \bgroup \em et al.\egroup }{2023}]{agentverse2024}
Weize Chen, Yusheng Su, Jingwei Zuo, et~al.
\newblock Agentverse: Facilitating multi-agent collaboration and exploring emergent behaviors, 2023.

\bibitem[\protect\citeauthoryear{Chen \bgroup \em et al.\egroup }{2025}]{critical2025}
Jianming Chen, Yawen Wang, Junjie Wang, et~al.
\newblock Understanding individual agent importance in multi-agent system via counterfactual reasoning.
\newblock {\em Proceedings of the AAAI Conference on Artificial Intelligence}, 39(15):15785–15794, April 2025.

\bibitem[\protect\citeauthoryear{Duan \bgroup \em et al.\egroup }{2026}]{uprop2025}
Jinhao Duan, James Diffenderfer, et~al.
\newblock {UP}rop: Investigating the uncertainty propagation of {LLM}s in multi-step decision-making, 2026.

\bibitem[\protect\citeauthoryear{Gan \bgroup \em et al.\egroup }{2025}]{snowball2025}
Zeyu Gan, Yun Liao, and Yong Liu.
\newblock Rethinking external slow-thinking: From snowball errors to probability of correct reasoning.
\newblock In {\em Forty-second International Conference on Machine Learning}, 2025.

\bibitem[\protect\citeauthoryear{Gao \bgroup \em et al.\egroup }{2023}]{gsmhard2025}
Luyu Gao, Aman Madaan, Shuyan Zhou, et~al.
\newblock Pal: Program-aided language models, 2023.

\bibitem[\protect\citeauthoryear{Goyal \bgroup \em et al.\egroup }{2024}]{llmguard2024}
Shubh Goyal, Medha Hira, Shubham Mishra, et~al.
\newblock Llmguard: Guarding against unsafe {LLM} behavior.
\newblock {\em CoRR}, abs/2403.00826, 2024.

\bibitem[\protect\citeauthoryear{Grattafiori \bgroup \em et al.\egroup }{2024}]{llama3_2024}
Aaron Grattafiori, Abhimanyu Dubey, Abhinav Jauhri, Abhinav Pandey, et~al.
\newblock The llama 3 herd of models, 2024.

\bibitem[\protect\citeauthoryear{Grunde{-}McLaughlin \bgroup \em et al.\egroup }{2025}]{llmchains2024}
Madeleine Grunde{-}McLaughlin, Michelle~S. Lam, Ranjay Krishna, et~al.
\newblock Designing {LLM} chains by adapting techniques from crowdsourcing workflows.
\newblock {\em {ACM} Trans. Comput. Hum. Interact.}, 32(3):27:1--27:57, 2025.

\bibitem[\protect\citeauthoryear{Islam \bgroup \em et al.\egroup }{2024}]{mapcoder2024}
Md.~Ashraful Islam, Mohammed~Eunus Ali, and Md~Rizwan Parvez.
\newblock Mapcoder: Multi-agent code generation for competitive problem solving, 2024.

\bibitem[\protect\citeauthoryear{Kwon \bgroup \em et al.\egroup }{2023}]{vllm2023}
Woosuk Kwon, Zhuohan Li, Siyuan Zhuang, et~al.
\newblock Efficient memory management for large language model serving with pagedattention.
\newblock In {\em Proceedings of the ACM SIGOPS 29th Symposium on Operating Systems Principles}, 2023.

\bibitem[\protect\citeauthoryear{Li \bgroup \em et al.\egroup }{2025}]{behaviorchain2025}
Rui Li, Heming Xia, Xinfeng Yuan, et~al.
\newblock How far are llms from being our digital twins? a benchmark for persona-based behavior chain simulation, 2025.

\bibitem[\protect\citeauthoryear{Madaan \bgroup \em et al.\egroup }{2023}]{selfrefine2023}
Aman Madaan, Niket Tandon, Prakhar Gupta, et~al.
\newblock Self-refine: Iterative refinement with self-feedback.
\newblock In {\em Thirty-seventh Conference on Neural Information Processing Systems}, 2023.

\bibitem[\protect\citeauthoryear{Manakul \bgroup \em et al.\egroup }{2023}]{selfcheckgpt2023}
Potsawee Manakul, Adian Liusie, and Mark Gales.
\newblock Selfcheck{GPT}: Zero-resource black-box hallucination detection for generative large language models.
\newblock In {\em The 2023 Conference on Empirical Methods in Natural Language Processing}, 2023.

\bibitem[\protect\citeauthoryear{Peng \bgroup \em et al.\egroup }{2025}]{peng-etal-2025-stepwise}
Jingyu Peng, Maolin Wang, Xiangyu Zhao, et~al.
\newblock Stepwise reasoning error disruption attack of llms, 2025.

\bibitem[\protect\citeauthoryear{Qian \bgroup \em et al.\egroup }{2025}]{qian2025scaling}
Chen Qian, Zihao Xie, YiFei Wang, et~al.
\newblock Scaling large language model-based multi-agent collaboration.
\newblock In {\em The Thirteenth International Conference on Learning Representations}, 2025.

\bibitem[\protect\citeauthoryear{tse Huang \bgroup \em et al.\egroup }{2025}]{faultyagents2025}
Jen tse Huang, Jiaxu Zhou, et~al.
\newblock On the resilience of llm-based multi-agent collaboration with faulty agents, 2025.

\bibitem[\protect\citeauthoryear{Wang \bgroup \em et al.\egroup }{2024}]{mmlupro2024}
Yubo Wang, Xueguang Ma, Ge~Zhang, et~al.
\newblock {MMLU}-pro: A more robust and challenging multi-task language understanding benchmark.
\newblock In {\em The Thirty-eight Conference on Neural Information Processing Systems Datasets and Benchmarks Track}, 2024.

\bibitem[\protect\citeauthoryear{Wei \bgroup \em et al.\egroup }{2023}]{cot2022}
Jason Wei, Xuezhi Wang, Dale Schuurmans, et~al.
\newblock Chain-of-thought prompting elicits reasoning in large language models, 2023.

\bibitem[\protect\citeauthoryear{Wu \bgroup \em et al.\egroup }{2024}]{autogen2023}
Qingyun Wu, Gagan Bansal, Jieyu Zhang, et~al.
\newblock Autogen: Enabling next-gen {LLM} applications via multi-agent conversations.
\newblock In {\em First Conference on Language Modeling}, 2024.

\bibitem[\protect\citeauthoryear{Wu}{2023}]{wu2023autogen}
Qingyun Wu.
\newblock Autogen: Enabling next-gen llm applications via multi-agent conversation.
\newblock {\em arXiv preprint}, arXiv:2308.08155, 2023.
\newblock ICLR 2024 submission.

\bibitem[\protect\citeauthoryear{Yang \bgroup \em et al.\egroup }{2025}]{qwen3report2025}
An~Yang, Anfeng Li, Baosong Yang, et~al.
\newblock Qwen3 technical report, 2025.

\bibitem[\protect\citeauthoryear{Yao \bgroup \em et al.\egroup }{2023}]{react2023}
Shunyu Yao, Jeffrey Zhao, Dian Yu, et~al.
\newblock React: Synergizing reasoning and acting in language models.
\newblock In {\em The Eleventh International Conference on Learning Representations}, 2023.

\bibitem[\protect\citeauthoryear{Zhang \bgroup \em et al.\egroup }{2025}]{aflow2025}
Jiayi Zhang, Jinyu Xiang, Zhaoyang Yu, et~al.
\newblock {AF}low: Automating agentic workflow generation.
\newblock In {\em The Thirteenth International Conference on Learning Representations}, 2025.

\bibitem[\protect\citeauthoryear{Zou \bgroup \em et al.\egroup }{2024}]{zou2024circuit}
Andy Zou, Long Phan, et~al.
\newblock Improving alignment and robustness with circuit breakers, 2024.

\end{thebibliography}

\end{document}